\documentclass[aps,twocolumn,prl,showpacs,color,psfig,epsf]{revtex4}
\usepackage{amsmath}
\usepackage{color}
\usepackage{epsf}
\usepackage{graphicx}
\begin{document}

\title{Nonequilibrium phase transition in the sedimentation of
reproducing particles}

\author{C. Barrett-Freeman,  M. R. Evans, D. Marenduzzo,
W. C. K. Poon}
\affiliation{SUPA and School of Physics, The University of Edinburgh, Mayfield Road, Edinburgh, EH9 3JZ, U.K.}

\begin{abstract}
We study numerically and analytically the dynamics of a sedimenting suspension 
of active, reproducing particles, such as growing bacteria in a gravitational 
field. In steady state we find a non-equilibrium phase transition between a 
`sedimentation' regime, analogous to the sedimentation equilibrium of passive 
colloids, and a `uniform' regime, in which the particle density is constant in 
all but the top and bottom of the sample. We discuss the importance of 
fluctuations in particle density in locating the phase transition point, 
and report the kinetics of sedimentation at early times.  
\end{abstract}


\maketitle

About a century ago, Einstein showed theoretically and Perrin
demonstrated experimentally that in a dilute colloidal suspension, the
particle density, $n$, as a function of height, $z$, is given by the
barometric distribution: $n(z) = n(0)\exp(-z/z_0)$, where $z_0$ is the
sedimentation height \cite{Perrin}. This distribution results from a
subtle interplay between thermal diffusion, hydrodynamics and
gravity. Diffusion and hydrodynamics are related via the
Stokes-Einstein formula, a form of the fluctuation-dissipation theorem
for equilibrium systems, while balancing gravity and Brownian motion
gives $z_0 = D/v_s$, where $D$ is the particles' diffusion coefficient
and $v_s$ their sedimentation speed. The barometric distribution
applies when the suspension is so dilute that interparticle potential
interactions (excluded volume, Coulomb, etc.) can be neglected. Modern
colloid physics has focussed on the behavior of {\it concentrated}
suspensions \cite{Pusey91}.

An equally interesting avenue to explore is that of {\it active}
particles (APs)\cite{AP}. Specifically, we consider APs able to propel
themselves in such a way that their long-time motion is diffusive,
i.e. each particle's mean-squared displacement from an initial
position, $\langle r^2(t) \rangle$, increases linearly with time, $t$,
so that (in three dimensions) $\langle r^2 \rangle = 6D_{\rm eff} t$,
where $D_{\rm eff}$ is an effective diffusion coefficient. The
swim-tumble-swim motion of an {\em Escherichia coli} bacterium ($\sim
2 \mu$m$\times 1 \mu$m spheroclinder, average density $\rho_b =
1.08$~g/cm$^3$) is an example \cite{Berg03a}, for which experiments
give $D_{\rm eff} \sim {\cal O}(10^2 \mu$m$^2$s$^{-1}$)
\cite{BergTurner90}. An equivalent passive colloid has $D \sim 0.5$
$\mu$m$^2$s$^{-1}$ at 300 K, so that an {\em E. coli} functions at an
effective temperature of ${\cal O}(10^4$~K): it is far from equilibrium. 
Mimicking bacteria, we assume our APs can also `reproduce' and 'die': their
number density may therefore change with time. Both motility and reproduction
require energy intake, although this is nowhere explicit in what
follows.

In this Letter, we study the behavior of (effectively) diffusing and
reproducing, non-interacting APs {\em in a gravitational field}. This
may model, for instance, a dilute suspension of motile {\em E. coli}
that are growing but not responding to chemical gradients
(i.e. non-chemotactic). Such a system is, arguably, the simplest
example of `active soft matter'. Does this paradigmatic AP system
differ significantly from its passive counterpart (dilute colloidal
sedimentation equilibrium)? And if so, how?

We first report 
{stochastic} simulations of the dynamics of diffusing and
reproducing APs in a gravitational field. We then interpret these
results by analysing a continuum equation describing the evolution of
a density profile of a dilute sedimenting AP fluid.  We find the
steady state profile, as well as the dynamic pathway leading to it.
Even a mean-field description reveals a much richer phenomenology for
APs than passive colloids. We find a non-equilibrium phase transition
between a `sedimentation' regime with exponential profile, and another
regime showing essentially constant density in the bulk of the
suspension. Using realistic parameter values, we predict that one may
switch between the two phases by modifying the growth rate of a real
system of bacteria.  Close to this transition, there exist novel
'sedimentation bands' in which a region of uniform AP density coexists
with an AP-depleted region. This may usefully be compared to the
phenomenon of shear banding.
We also analyse the role of noise, and show that its presence  
shifts the transition point. Finally, we discuss generalisations of
our equations, and how our results may relate to real bacterial
suspensions.

We use a stochastic algorithm to simulate the coupled biased diffusion
and reproduction/death of APs. We consider a column of sedimenting
APs as a discrete lattice of sites $i = 1,\dots L$ with the number of
AP occupying each lattice site specified as $n_i(t)$.  Gravity acts
towards $i=0$.  At each time step $t \to t+ \Delta t$ the array of
occupation numbers is updated according to a `multiply' step or a
`move' step chosen with probabilities $w/(1+w)$ or $1/(1+w)$
respectively where $w = \alpha \left[ 1+ \sum_{i=1}^L n_i^2/(\rho_0
\sum_i n_i(t))\right]$ 
{is the ratio of the total rate of reproduction/death per particle to the  total rate of moving
per particle,}
and $\alpha$, $\rho_0$ are parameters to be
discussed below~\cite{note_nonlocal}.
  
In a `move' update each AP moves independently up or
down with probability $p$ or $1-p$. In a `multiply' update, at each
site $i$ each particle is replaced by two particles with probability
$\rho_0/(\rho_0+ n_i(t))$ or removed with probability $n_i(t)/(\rho_0+
n_i(t))$.  We impose no flux, or reflecting, boundary conditions at
the top and the bottom of the container ($i=0$ and $i=L$).

If $\Delta z$, $\Delta t$ represent respectively the spatial
and temporal steps, 
then the continuum limit of our Markov process leads to a
diffusion constant $D = \frac{\Delta z^2}{2 \Delta t}$ and
sedimentation velocity $v = \left(1-2p\right)\frac{\Delta z} {\Delta
t}$.  The parameter $p$ controls the strength of the gravitational
force; 
$\alpha$ controls the rate of reproduction/death and $\rho_0$
gives the  value of the occupation in which reproduction
and death are balanced. In a real bacterial suspension, $\alpha$ will
be the growth rate (medium dependent, but $\gtrsim {\rm hour}^{-1}$
for {\em E. coli}) and $\rho_0$ is the saturation cell density
($\sim10^{9}$~cells/cm$^3$ for {\em E. coli}\cite{Bailey86}).

Note that the state where the lattice is devoid of APs is an absorbing
state of the dynamics in the algorithm, which we call model I.  We
also considered a variation of the dynamics, which we refer to as
model II, in which a depopulated site is allowed to be repopulated
spontaneously (during the reproduction/death move).

Our simulations suggest that there is a phase transition between two
different regimes. If the APs reproduce slowly ($\alpha$ small), we
obtain a steady-state density profile which decays sharply with $z$
(Fig. 1a).  We call this the `sedimentation regime' in analogy with
the similar behaviour of passive colloids under gravity. If the growth
rate exceeds a threshold, $\alpha_c$, the steady state is one
with a uniform density throughout the bulk of the sample, with
depleted and enriched layers in the top and the bottom of the
container (Fig. 1b). In model II, we find that the critical value is
close to $\alpha_c = \frac{v^2}{4D}$; whereas in model I, the presence
of an absorbing state widens  the sedimentation regime, shifting
$\alpha_c$ to slightly larger values.

To understand the simulations, we take a continuum approach and coarse
grain our discrete model by defining a 1-dimensional profile of AP
density, $\rho(z,t)$ (gravity acts along the negative $z$ axis).  To
make contact with the numerical simulations, and on general grounds,
we may assume that the density profile obeys the following dynamic
equation of motion:
\begin{equation}
\frac{\partial \rho}{\partial t} = D \frac{\partial^2 \rho}{\partial z^2}+
v \frac{\partial \rho}{\partial z}+\alpha\rho 
\left(1-\frac{\rho}{\rho_0}\right)+\Gamma(\rho)\eta(t).
\label{sedimentation_1d}
\end{equation}
The first term  represents diffusion, the second term represents the 
sedimentation due to gravity and
the third term represents reproduction/death.
Here, $D$ and $v$ have the same meaning as in the numerical
simulations.  An important dimensionless control parameter is
$\theta=v/\sqrt{D \alpha}$.  The last term in (\ref{sedimentation_1d})
represents noise, and $\eta(t)$ is a white noise with unit variance,
while $\Gamma(\rho)$ is a function to be specified. We first
consider the deterministic case, $\Gamma(\rho)=0$ which 
reduces to a mean-field description.

\begin{figure}
\centerline{\includegraphics[width=8.cm]{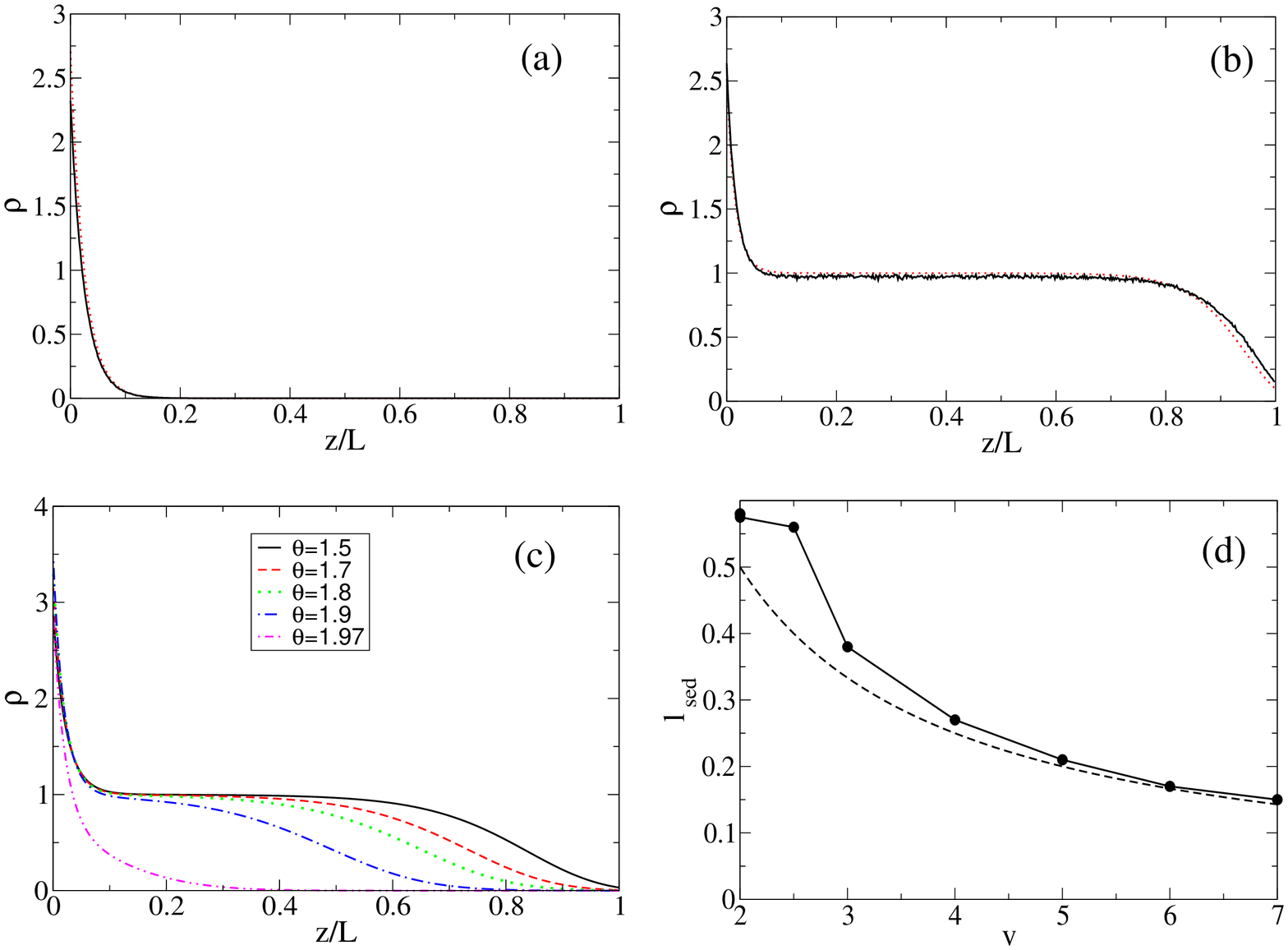}}
\caption{The average density, $\rho$, versus height, $z$, for a suspension of motile and reproducing active particles in a gravitational field. Simulation results (model I), with (a) $\theta=3.6$, and (b) $\theta=1.4$.
In (c) we show different numerical solutions of the noiseless continuum 
equation Eq. \ref{sedimentation_1d}, while (d) shows the dependency
of the sedimentation length, on the velocity (in units of $\sqrt{D\alpha}$; the
dashed line gives the colloidal value at $\alpha=0$, which is denoted
by $z_0$ in the text).} 
\end{figure}

Considering the steady state  ($t\to\infty$),
if $\alpha=0$, we get back passive colloid sedimentation equilibrium
$\rho(z)\propto \exp(-vz/D)$ .
However, for $\alpha >0$ one obtains a nonlinear equation for which an exact solution is not available.
Therefore we  perform a perturbation expansion for small $\alpha$ using
the first 30 terms of a series expansion $\rho(z)=\sum_n \alpha^n
\rho^{(n)}(z)$. The series converges for
$\alpha<\alpha_c=\frac{v^2}{4D}$ and diverges otherwise. This
corroborates the simulation results and suggests that indeed a phase
transition occurs as the $\theta=v/\sqrt{D \alpha}$ goes through 2.

Eq. \ref{sedimentation_1d} with $v=0$ and no noise is the well-known
Fisher-KPP equation and admits advancing waves with velocity 
$v_w=2\sqrt{D\alpha}$ as solutions
\cite{vansaarlos}, we can understand this phase transition as a
competition between gravity forcing the bacteria downwards with
velocity $v$ and a travelling wave of proliferation which advances
upward. For $v>v_w$, the sedimentation wave wins over the Fisher wave
and leads to an exponential profile, while for $v<v_w$ the Fisher wave
leads to a uniform bacterial density throughout the sample. 
{However, Eq. \ref{sedimentation_1d} actually yields 
a transient travelling wave for $v>v_w$ only.}
{The transition is reminiscent of transitions in interface 
depinning \cite{MK01}; in a branching random walk with an absorbing wall 
\cite{DS07}, and of extinction transitions in inhomogeneous biological
systems \cite{DSN00}.}
We note that a linearised version of our Eq. (\ref{sedimentation_1d}) with
$\Gamma=0$ and different boundary conditions was considered in Ref. 
\cite{speirs} to describe microorganisms advected in a river and 
resultant extinction. We believe that our analysis should
apply to this problem as well. 
{We also stress that the
transition would be washed away by translational invariance or
periodic boundary conditions, which are usually considered 
\cite{vansaarlos,DSN00}.}

The existence of a non-equilibrium phase transition at
$\theta=\theta_c=2$ is confirmed by numerical solution of the
noiseless version of Eq. \ref{sedimentation_1d} using a standard
finite difference scheme.  An example of a series of steady state
solutions for different values of $\theta$ is shown for a sample of
size $L=20 \sqrt{D/\alpha}$, in Fig. 1c. Increasing the system size,
the segment of the sample at $\rho=1$ in the uniform regime increases
(data not shown), analogous to equilibrium phase transitions. It is
interesting to consider the behavior of the decay length of the
exponential density profile in the sedimentation regime (an effective
sedimentation length) as a function of $|\theta - \theta_c|$. This
sedimentation length is akin to a scaling length in an equilibrium
phase transition. We find that after correcting for a small
$L$-dependent shift in $\theta_c$, the sedimentation length does {\em
not} diverge at the transition, and is only at most $\sim 20\%$ larger
than the corresponding sedimentation length with no growth ($\alpha =
0$).  If we focus on the steady
state concentration value e.g. in the middle of the sample, it
switches abruptly, for $L \rightarrow \infty$, from 0, for
$\theta>\theta_c$, to $1$, for $\theta<\theta_c$.
These obervations are consistent with a discontinous  phase
transition.

For large but finite systems, we find evidence of an intriguing
spontaneous banding of the sedimenting APs, 
which occurs very close to
$\theta = \theta_c$.  An example is shown in Fig. 2, in which two
steady-state regions coexist in the bulk of the sample, one in which
$\rho$ is practically 0 and another one in which it is $\sim 1$. This
kink-like solution, which we call a `sedimentation band', is similar
to a {\em cline}, found in the population biology literature when
solving a reaction-diffusion equation, similar to
Eq. \ref{sedimentation_1d}, but with $v=0$ and a reaction term which
is cubic in $\rho$ \cite{barton}.  In our case, however, the existence
of sedimentation bands is more surprising, as without advection the
state at $\rho=0$ is unstable, and bands arise due to the vicinity of
a phase transition. In this respect, our sedimentation banding is more
akin to shear banding, which is obtained when some complex fluids such
as liquid crystals and worm-like micelles in the isotropic or
disordered phase, are subjected to a shear, slightly smaller than that
needed to order them completely \cite{shear_banding}.
Sedimentation bands appear in an increasingly small window of $\theta$
as $L$ increases, and disappear in the thermodynamic ($L\to \infty$) limit.

\begin{figure}
\centerline{\includegraphics[width=7.cm]{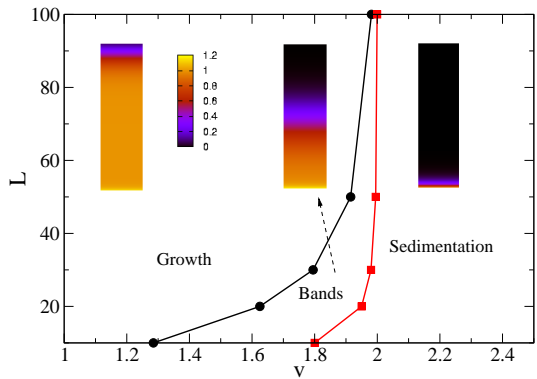}}
\caption{Steady state diagram in the $(v,L)$ plane for the 
noise-less version of Eq. \ref{sedimentation_1d}. A density profile is
classified as `banded' if the cline stays in the bottom $75\%$ of the sample. 
Typical concentration profiles in the various
regimes are also shown, together with a color scale for the density. }
\end{figure}

\begin{figure}
\includegraphics[width=9.cm]{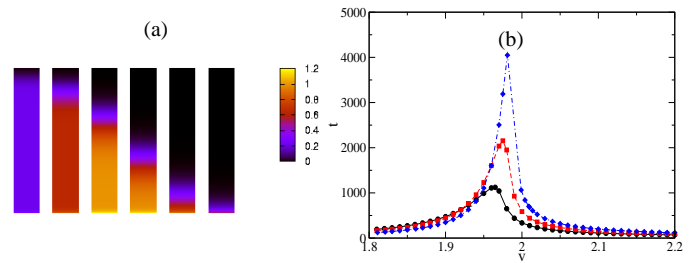}
\caption{(a) Dynamics of the density profile in
a system with $L=30$ (in units of $\sqrt{D/\alpha}$), 
and with $\theta=2.1$. The times corresponding to each profile are
(from left to right) 0.99, 2.8, 17.5, 62.2, 121.4, 250 (in units of 
$\alpha^{-1}$, the bottom 5\% of the sample is cut for colour readability).
(b) Plots of the time needed to get to steady state (in units
of ${\alpha}^{-1}$), as a function of $v$ (in units of $\sqrt{D\alpha}$),
with a system with $L=20$ (solid line), $L=28$ (dashed line)
and  $L=40$ (dot-dashed line). ($L$
is in units of  $\sqrt{D/\alpha}$.)}
\end{figure}

Would it be possible to observe the transition we predict in a real
bacterial suspension? For {\em E. coli} in water, $v \lesssim 0.1
\mu$ms$^{-1}$ \cite{Klaus97} and is fixed, while $D \gtrsim 10^2
\mu$ms$^{-1}$ \cite{BergTurner90}. In rich, well aerated media
maintained at the optimal temperature of 37$^{\circ}$C, the population
doubles every $\sim 20$~minutes, giving $\alpha \sim
10^{-3}$~s$^{-1}$, and $\theta \approx 0.3$. It is possible to culture
the bacterium in what is known as `motility buffer', in which $D$ is
maintained, but growth essentially stops ($\alpha \rightarrow 0$),
allowing the tuning of $\theta$ from 0.3 through 2 to arbitrarily
large values, thus permitting the observation of our transition in
principle.

In this context, it is important to note that the pathway to steady
state may be quite slow. This is particularly true close to
$\theta_c(L)$, and in the region where sedimentation bands form. In
the banding regime for large sample size $L$, we also find that the
behaviour of the part of the sample close to the top, or just after
the boundary of the band, display non-monotonic behavior.  The density
first increases, as if the systems transiently entered the uniform
regime, to decay later on to reach equilibrium (see Fig. 3a). The time
scale needed to reach equilibrium, $t_{\rm eq}$, is plotted in Fig. 3b
as a function of the distance from the transition point. Larger
systems take longer to equilibrate, while close to criticality we find
that $t_{\rm eq}$ increases as a power law of
$\left|{\theta-\theta_c}\right|^{-a}$, with $a\simeq 1$ above the transition, 
consistent with \cite{DS07} and confirming the presence of a phase-transition at
$\theta_c=2$. These results applied to APs with $\alpha=10^{-4}$
s$^{-1}$ predict that close to the transition, it may take up to
several months for a column of ~10 cm height to reach steady state.

Next, we discuss the role of noise, i.e. the case $\Gamma(\rho)\ne 0$
in Eq.~\ref{sedimentation_1d}.  If $\rho=0$ is to be an absorbing
state, we need to go beyond the case $\Gamma(\rho)=\Gamma_0>0$,
and consider instead a density-dependent amplitude $\Gamma$. 
The natural choice is $\Gamma=\Gamma_0\sqrt{\rho}$, 
which is justified by a central limit argument relating the variance of the noise
to the number of active particles\cite{HH,noise_note}. 
Increasing
$\Gamma_0$ favours large fluctuations and may locally bias the system
towards the absorbing state.
We observe that  as $\Gamma_0$
is increased the sedimentation regime is enhanced  at the expense of the growth one
(see Fig. 4). This is consistent with  a negative shift in the Fisher wave velocity
due to noise \cite{BD97}.
{The 
non-equilibrium phase boundary shown in Fig. 4 is found by locating
the maxima of the order parameter fluctuations.} 

Eq. \ref{sedimentation_1d} 
with $v=0$ describes directed
percolation (DP) \cite{HH}, the generic university class of nonequilibrium transitions 
from an absorbing state ($\rho=0$) to a fluctuating  state  ($\rho >0$).
The dimensionless control parameter  is in this case
$\Gamma_0/\left({D\alpha}\right)^{1/4}$. Thus Fig. 4 becomes
an extension of the DP phase diagram to include a $v$ axis. Intriguingly
the transition we have studied for  $\Gamma_0$ is discontinuous whereas the DP transition
at $v=0$ is continuous. Therefore one may speculate that there is a 
singular or even tricritical 
point along the 
critical curve $v_c(\Gamma_0)$. 
{This scenario might be 
similar to what occurs in the XY model, where switching the XY 
spin into a velocity can result in a discontinuous transition \cite{GC04}.}

\begin{figure}
\includegraphics[width=5.3cm]{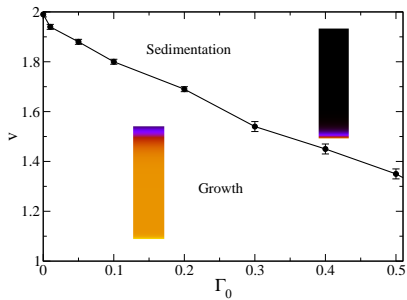}
\caption{The dependence of the critical velocity, $v_c$
(in units of $\sqrt{D\alpha}$) on 
$\Gamma_0$ (in units of $\left(D\alpha\right)^{1/4}$), for
$L=200$ $\sqrt{D/\alpha}$. }
\end{figure}

We have already pointed out that with typical values of $v$ and $D$ in
a suspension of {\em E. coli}, it may be possible to `tune' the growth
constant $\alpha$ to bring the system from the sedimentation to the
uniform regime according to our predictions. In reality, these
experiments need stringent controls, e.g. to make sure that the
bacteria are not engaged in any form of chemotaxis, which would render
$D$ dependent on concentrations of chemical species (nutrient, oxygen,
waste products, \ldots). The extra level of complexity introduced can
be modelled by adding a chemotactic term to
Eq. \ref{sedimentation_1d}, and coupling it to a reaction-diffusion
equation, e.g. as in the Keller-Segel model \cite{keller}. Moreover,
we have shown that the time scales for reaching steady state can be
long, reaching ${\cal O}(10^3)$ in units of the inverse growth rate,
$\alpha^{-1}$, in the vicinity of the transition. But in a bacterial
culture, $\alpha$ itself is only approximately constant during what is
known as the `exponential' growth phase, after which saturation in
population density and then death follow. Thus, steady-state
experiments at $\theta \approx \theta_c$ are likely impractical.

Clearly, our model also neglects hydrodynamic interactions  \cite{hydro}, which may
(for example) cause swimmers to attract. 
Thus, hydrodynamics could have highly non-trivial
effects, e.g. concerning the approach to steady states. However, we
believe that the qualitative features of the transition we have
identified may survive, because these are due to the competition
between gravity and Fisher wave fronts, which should be generic.
{Also, there are non-swimming bacteria,
such as e.g. {\it Sinorhizobium Melitoti}, where hydrodynamics is less 
important.}

To summarize, we have shown that the physics of motile and
`reproducing' active particles in a gravitational field yields a
surprisingly non-trivial phenomenology.  We have found, by simulations
and analytic work, that increasing the growth rate from zero, the
system makes an abrupt transition from a sedimentation regime in which
the density decays exponentially with the distance from the bottom of
the container, to a uniform growth-dominated regime in which the
density is practically constant spatially except for two boundary
layers at the top and at the bottom to satisfy boundary
conditions. The essential physics is that of a balancing between a
downward gravitational flux, and an upward diffusion-growth
flux. Using values appropriate to {\em E. coli}, we predict that it
may be possible to observe this transition in a real bacterial
suspension.

We thank O. E. Croze and B. Derrida for helpful discussions.


\begin{thebibliography}{99}
\bibitem{note_nonlocal} 
{The term $w$ is non-local as we average
the rate of reproduction $\alpha$ and the death rate $\alpha n_i/\rho_0$
over all particles.}
\bibitem{noise_note}Note that we disregard possible 
additional conservative noise terms as irrelevant, as in directed percolation.
\end{thebibliography}
\end{document}